\documentclass[twocolumn,reprint,aps,prb,floatfix,longbibliography]{revtex4-2}
\usepackage[T1]{fontenc}
\usepackage{amsfonts}
\usepackage{comment}
\usepackage{amsmath,amssymb,epsfig,color}
\usepackage{float}
\usepackage{amsmath}
\usepackage{amssymb}
\usepackage{tabularx}
\usepackage{makecell}
\usepackage{booktabs}
\usepackage{url}
\usepackage{graphicx}
\usepackage{dcolumn}
\usepackage{bbold}
\usepackage{physics}
\usepackage{multirow}
\usepackage{pifont}
\usepackage{siunitx}
\usepackage[normalem]{ulem}

\usepackage{standalone}

\usepackage{hyperref}
\newcommand{\Lvec}{\mathbf L}
\newcommand{\kvec}{\mathbf k}

\newcommand{\Evec}{\mathbf E}

\newcommand{\nhat}{\hat{\mathbf n}}

\begin{document}

\title{
Deterministic Electrical Switching in Altermagnets via Surface Antisymmetry Groups
}

\author{K. D. Belashchenko}
\affiliation{Department of Physics and Astronomy and Nebraska Center for Materials and Nanoscience, University of Nebraska-Lincoln, Lincoln, Nebraska 68588, USA}
\date{\today}

\begin{abstract}
A surface antisymmetry group framework is developed to establish design rules for deterministic electrical switching of the N\'eel vector in a film of a collinear bipartite antiferromagnet. In centrosymmetric altermagnets, where linear current-induced spin-orbit torques vanish in the bulk, staggered effective fields can nevertheless exist as a macroscopic interfacial response, whose allowed tensor form is determined by the surface antisymmetry point group for the given surface orientation. Such effective fields do not require sublattice equivalence to be broken by the surface and can arise via the Edelstein mechanism driven by sublattice-odd interfacial spin-orbit coupling. General symmetry obstructions to bipolar switching that are valid to all orders in torque angular harmonics are also derived.
Separately, the structure of the spin conductivity tensor determines which surface orientations allow transverse spin current generation via the nonrelativistic spin-splitter effect. Taken together, these symmetry-enforced properties establish which surface orientations of $d$-wave altermagnets can serve as deterministically switchable spin current sources in spin-torque heterostructures. Because the design rules are based solely on the surface antisymmetry point group, the symmetry-allowed staggered effective fields are robust against averaging over equilibrium surface roughness.
\end{abstract}

\maketitle

\section{Introduction}

Altermagnets are collinear magnets whose electronic structure exhibits momentum-dependent spin splitting in the nonrelativistic limit despite vanishing net magnetization (see reviews \cite{Smejkal2022a,Mazin2022,Bai2024,Tamang2024,Song_review_2025} and references therein). Altermagnets are of significant interest for spintronic applications because their symmetry generally allows the detection of the N\'eel vector via anomalous Hall effect \cite{Smejkal2020} and generation of spin-polarized currents via the spin-splitter effect \cite{SpinSplitting-Gonzalez,Junwei2021,SpinSplitting-Bai,SpinSplitting-Karube}. The latter enables the use of altermagnets as efficient spin current sources with perpendicular spin polarization in spin-torque devices \cite{Bose2024,Song_review_2025}. An essential feature for device applications is the ability to initialize the N\'eel vector $\Lvec$ deterministically by electrical means. With such initialization, altermagnetic layers could both store information and control other layers via spin currents.

The majority of known altermagnets are centrosymmetric. An inversion center in an altermagnet necessarily maps each magnetic sublattice onto itself; otherwise it would enter as the $PT$ operation and forbid altermagnetism. If the inversion centers are located on the magnetic atoms, all linear current-induced spin torques are forbidden; more generally, the macroscopic spin torque averaged over a given sublattice vanishes in linear response.
Therefore, the initialization mechanism utilizing staggered current-induced effective fields (so-called N\'eel-type spin-orbit torque; NSOT \cite{Zelezny2014,Wadley2016,Gomonay2016,Zelezny2017,Watanabe2018,Manchon-RMP}) is unavailable in the bulk without breaking that inversion symmetry \cite{Chien2025RashbaEngineering}.

Nevertheless, deterministic field-free 180$^\circ$ switching was demonstrated in a monoclinically distorted CrSb ($11\bar20$) film, where it was attributed to a staggered effective field generated from the externally injected spin polarization by Dzyaloshinskii-Moriya interaction \cite{Zhou2025}. The present paper aims to establish when deterministic electrical (i.e., magnetic field-free) 180$^\circ$ switching of $\Lvec$ in an altermagnetic film is allowed by symmetry. The scope of this consideration excludes field-assisted or nondeterministic switching strategies. For example, deterministic 180$^\circ$ switching was demonstrated in Mn$_5$Si$_3$ using spin-orbit torque from an adjacent Pt layer \cite{Han2024Electrical180switching}, but it used an external magnetic field to introduce an asymmetry between the energy barriers separating the opposite N\'eel states.
Current-driven 90$^\circ$ reorientation of the N\'eel vector is also possible \cite{Song2025RuO290degrees}, but it does not distinguish between time-reversed partners $\Lvec$ and $-\Lvec$, leaving the sign of the anomalous Hall and spin-splitter effects undetermined. 
Nonlinear staggered current-induced torques \cite{Cao2026} are also beyond the scope of the present paper.
In the following, unless otherwise noted, \emph{switching} means deterministic, bipolar (i.e., reversible by current polarity) electrical initialization of a 180$^\circ$-bistable $\Lvec$ driven by torques linear in $\Evec$.

It was proposed that interfaces can enable sublattice-odd current-induced spin accumulations and facilitate deterministic switching into a specific $\Lvec$ domain \cite{Zhang2026DeterministicSwitching,sarkar2025deterministicswitchingaltermagnetsasymmetric}; this was illustrated by first-principles calculations for a specially designed film structure with broken inversion symmetry \cite{Chen2025ElectricalSwitchingAltermagnetism}. 
However, a general framework identifying the conditions for such control based on interface-induced symmetry lowering has been lacking. In particular, it has remained unclear when such sublattice-odd responses are robust against surface roughness rather than requiring precisely engineered surface terminations. Here it is shown that the presence of sublattice-odd axial current-induced responses is governed by the surface antisymmetry point group (SAPG), i.e., the subgroup of the bulk antisymmetry point group that leaves the surface normal $\nhat$ invariant. Based on this principle, symmetry design rules are derived to classify which surface orientations permit deterministic 180$^\circ$ electrical switching of the N\'eel order parameter in centrosymmetric altermagnets. The criterion of deterministic switchability adopted here is macroscopic; it determines which sublattice-odd responses survive after averaging over symmetry-related terminations and roughness without relying on any particular interfacial structure. 

\section{Problem setup}
\label{sec:setup}

Consider a single-crystalline altermagnetic film with a given surface normal $\nhat$. We will assume that the top and bottom interfaces are inequivalent, as is almost always the case in heterostructures, and that the statistical distribution of surface terminations is constrained only macroscopically by $\nhat$. Therefore, we postulate that this distribution is invariant under the surface point group, which is the subgroup of the bulk point group that leaves $\nhat$ invariant. It may, for example, correspond to thermodynamic equilibrium at the fabrication temperature. Thus, our consideration will describe macroscopic responses that are generically allowed for the given $\nhat$ and survive after averaging over interfacial roughness and equivalent termination domains \cite{Belashchenko2010,He2010,Lapa2020,Weber2024,belashchenko2026surfaceAM}. This approach to surface symmetry is different from considerations assuming a flat surface termination \cite{Lange2026,Sasioglu2026,Leeb2026,Hu2026}.

We further assume bipartite ordering such that all magnetic sites in the film may be uniquely assigned to sublattice A or sublattice B. This assumption ensures that the key antisymmetry operation (i.e., sublattice interchange) remains well defined throughout the film. 

We also assume the strong exchange limit, so that the two sublattices remain approximately antiparallel: $\mathbf{m}_\mathrm{A}\approx-\mathbf{m}_\mathrm{B}\approx\Lvec$.
Our primary focus will be on the fieldlike (FL) and dampinglike (DL) torques characterized by $\Lvec$-independent \emph{torque-polarization vectors} $\mathbf{p}^\alpha_\mathrm{FL}(\mathbf{E})$ and $\mathbf{p}^\alpha_\mathrm{DL}(\mathbf{E})$, where $\alpha$ denotes the sublattice:
\begin{align}
\mathbf{T}_\alpha(\mathbf{E}) = \mathbf{m}_\alpha \times \mathbf{p}^\alpha_{\mathrm{FL}}(\mathbf{E})
+\mathbf{m}_\alpha\times\bigl[
\mathbf{m}_\alpha\times\mathbf{p}^{\alpha}_{\mathrm{DL}}(\mathbf{E})\bigr],
\label{eq:torques}
\end{align}
and we interpret $\mathbf{p}^\alpha_\mu(\mathbf{E})$ ($\mu=\mathrm{FL}$ or DL) as macroscopic sublattice averages over the entire film. The case of a general angular dependence of the torque will be addressed separately below.

For a given in-plane electric field $\mathbf{E}$, symmetry restricts $\mathbf{p}^\alpha_\mu(\mathbf{E})$ to lie within a certain linear subspace of axial vectors determined by the interface symmetry. For example, in the standard, axially symmetric ferromagnet/nonmagnet spin-orbit torque bilayer with interface normal $\nhat$, this subspace is one-dimensional, with $\nhat \times \mathbf{E}$ providing the common polarization axis for both FL and DL torques \cite{Manchon-RMP}. In lower-symmetry situations the allowed axial subspace may have higher dimension, in which case FL and DL torque-polarization vectors need not share the same direction \cite{Amin2024}.

We focus on the case of bipolar switching between two time-reversed N\'eel domains ($\Lvec\leftrightarrow-\Lvec$). A writing current must initialize a given sign of $\Lvec$, independently of its initial state, and reversing the current polarity must select the opposite domain. In the case of a ferromagnet/nonmagnet bilayer, the DL and FL torques with polarization along $\nhat \times \mathbf{E}$ can both initialize the magnetization along this axis \cite{Manchon-RMP}. The situation is different in antiferromagnets where,
for torques in the form of Eq. (\ref{eq:torques}), a selection bias between $\Lvec$ and $-\Lvec$ requires at least one \emph{staggered} torque-polarization vector $\mathbf{p}^{(-)}_\mu=\mathbf{p}^\mathrm{A}_\mu-\mathbf{p}^\mathrm{B}_\mu$ to be finite \cite{Zelezny2017,Watanabe2018,Zhang2026DeterministicSwitching}, where the superscript $(-)$ designates that it is odd under sublattice interchange.

\section{Symmetry framework}

Linear response of a staggered axial vector $\mathbf{p}^{(-)}_\mu(\mathbf{E})$ to an in-plane electric field may be written as
\begin{equation}
p^{(-)}_{i} = \kappa^{(-)}_{ij} E_j ,
\end{equation}
where $\mathbf{E}\perp\nhat$, $\hat\kappa^{(-)}$ is a $3\times2$ axial-polar (fieldlike or dampinglike) torque-polarizability pseudotensor that is odd under sublattice interchange, and the subscript $\mu$ is dropped for clarity.
Nonvanishing elements of $\hat\kappa^{(-)}$ determine the components of $\Lvec$ that may be deterministically switched by the symmetry-allowed torque polarizations. In centrosymmetric altermagnets the bulk contribution to $\hat\kappa^{(-)}$ vanishes, and deterministic switching between $\Lvec$ and $-\Lvec$ must therefore rely on the reduced symmetry of the film. 

The form of a macroscopic response tensor is constrained by Neumann's principle, which requires it to be invariant under the symmetry operations of the crystal. For tensors that are odd under sublattice interchange, such as $\hat\kappa^{(-)}$, Neumann's principle must be applied to the antisymmetry point group \cite{Shubnikov1951}. To define antisymmetry \cite{Heesch1930}, the two sublattices are decorated by two labels (e.g., A and B); the antisymmetry operation is the label swap. A candidate element $g$ of the antisymmetry group combines a conventional spatial symmetry operation $R_g$ inherited from the crystallographic symmetry group and a label $\eta_g=\pm1$ denoting whether it includes an additional label swap. For the bulk material, the antisymmetry point group $G$ is identical to the nontrivial spin point group \cite{LITVIN1974538,Litvin:a14103,Smejkal1} that determines collinear altermagnetic order, with the opposite spins playing the roles of labels assigned to the two sublattices. Equivalently, antisymmetry and the staggered responses can be described using the one-dimensional irreducible representation of the order parameter \cite{McClarty2024,Watanabe2018}.

For an oriented film with surface normal $\nhat$, under our assumption of statistically invariant interface roughness, macroscopic response properties obey Neumann's principle with respect to the relevant surface point group \cite{Belashchenko2010}. The sublattice-odd tensor $\hat\kappa^{(-)}$ must therefore be invariant under the \emph{surface antisymmetry point group} $G_s(\nhat)=\{\,g\in G \mid R_g\hat{\mathbf n}=\hat{\mathbf n}\,\}$. Explicitly, for each element $g\in G_s(\nhat)$, the axial-polar tensor $\hat\kappa^{(-)}$ must satisfy
\begin{equation}
\hat\kappa^{(-)} = \eta_g (\det R_g) \, R_g \, \hat\kappa^{(-)} \, R_g^{-1},
\label{eq:symmetry_condition}
\end{equation}
and the allowed structure of $\hat\kappa^{(-)}$ may be obtained by symmetrizing a generic $3\times3$ tensor over the group $G_s(\nhat)$ and then extracting the components for $\mathbf{E}\perp\nhat$.

\section{Strong surface magnetization and the nonrelativistic limit}

Because altermagnets break macroscopic time-reversal symmetry ($T\tau$ is not a symmetry with any translation $\tau$, where $T$ is time reversal), a generic surface of an altermagnet carries a finite surface magnetization \cite{Andreev1996,Belashchenko2010}. Let us call the surface magnetization \emph{strong} if the equivalence of the two sublattices is broken by the surface. Strong surface magnetization (SSM) is allowed if the SAPG $G_s(\nhat)$ for the given surface orientation does not include any elements with $\eta_g=-1$ \cite{Belashchenko2010}.

SSM affects the absorption of the incident spin current. Consider the special case of an altermagnet (or, more generally, a $T\tau$-breaking antiferromagnet) with vanishing spin-orbit coupling that is interfaced with a nonmagnetic layer that injects spin current through that interface. Without spin-orbit coupling, the only available axial vector is the spin polarization $\mathbf{p}_{inj}$ of the injected spin current. Therefore, we must have $\mathbf{p}^{(-)}_\mu=\chi^{(-)}_\mu\mathbf{p}_{inj}$, where $\chi^{(-)}_\mu$ is a sublattice-odd scalar. The existence of such a scalar requires SSM; physically, if the sublattices are made inequivalent by the interface, they absorb the incident spin current unequally. This nonrelativistic mechanism has been previously discussed \cite{Zhang2026DeterministicSwitching}; however, as we will see below, it is forbidden for most high-symmetry surface orientations.

\section{Device considerations}

The above prescription can be applied to any bipartite antiferromagnet. Of particular interest for applications are centrosymmetric altermagnets belonging to the four $d$-wave spin point groups \cite{Smejkal1}. Such magnets generically exhibit the nonrelativistic spin-splitter effect \cite{SpinSplitting-Gonzalez,Junwei2021,SpinSplitting-Bai,SpinSplitting-Karube} and can generate  transverse spin current in heterostructures similar to spin-orbit-torque bilayers, ideally with a nonmagnetic spacer providing exchange decoupling between the altermagnet and the magnetic layer receiving the torque. Because the existence of a spin-splitter current $j^s_\perp$ flowing perpendicular to the interface also depends on the surface orientation, it is useful to consider it along with the structure of the $\hat\kappa^{(-)}$ tensor.

The most favorable case for applications is when $j^s_\perp$ generated in the altermagnet is polarized perpendicular to the interface ($\perp$-polarized), enabling the switching of perpendicular magnetization \cite{Zhu2023}. This desirable $\perp$-polarization is often called ``unconventional'' because it requires mirror-symmetry breaking in spin-orbit-torque bilayers \cite{Manchon-RMP}. Because the spin-splitter current is polarized parallel to $\Lvec$, direct initialization of a $\perp$-polarized altermagnetic spin current source requires $\mathbf{p}_\mu^{(-)}\cdot\nhat\neq0$.
Thus, an ideal altermagnetic spin-splitter layer satisfies the following conditions: (1) $\mathbf{p}^{(-)}_\mu\cdot\nhat\neq0$, (2) $j^s_\perp\neq0$, and (3) easy-axis anisotropy along $\nhat$, which may be facilitated by surface anisotropy or magnetoelastic coupling. The condition (1) is governed by SAPG, (2) is a bulk symmetry property projected onto the given $\nhat$, and (3) is material-specific.

\section{Symmetry design rules}

Table~\ref{tab:atlas} lists symmetry-allowed responses for all centrosymmetric altermagnetic bulk antisymmetry point groups with various surface orientations, including the structure of the $\hat\kappa^{(-)}$ tensor, the perpendicular nonrelativistic spin-splitter current $j^s_\perp$, and the presence of SSM. Note that the nonrelativistic spin-splitter effect is only allowed in $d$-wave altermagnets. 

\begin{table*}[ht]
\caption{\label{tab:atlas}
Properties of altermagnetic films belonging to centrosymmetric bulk antisymmetry point groups $G$ for different surface orientations $\nhat$.
$G_s(\nhat)$: SAPG. $\hat\kappa^{(-)}$: staggered FL or DL torque-polarizability tensor. The rows giving the perpendicular component  $p^{(-)}_\perp$ are highlighted in bold. W$_\perp$: at least one such component is present, indicating that deterministic switching of $L_\perp$ is symmetry-allowed. $j_\perp^s$: spin current flowing perpendicular to the film due to the nonrelativistic spin-splitter effect. $(E_1,E_2)$: basis for the in-plane electric field; primed coordinates correspond to a rotated reference frame. SSM: strong surface magnetization. The first four bulk antisymmetry groups $G$ are $d$-wave, and the rest are $g$-wave and $i$-wave. Other high-symmetry $\nhat$ directions for $g$-wave and $i$-wave cases already appear elsewhere in the table.
Zero elements of $\hat\kappa^{(-)}$ and W$_\perp=\mathrm{No}$ entries marked with a dagger hold only for torques of the form of Eq. (\ref{eq:torques}); others are enforced to all orders by a toggle or unipolar obstruction (Table \ref{tab:switching_obstructions}) for $\Evec$ along the corresponding basis vector. All W$_\perp=\mathrm{No}$ entries without a dagger are also protected for any field orientation by a ${}^12$ axis due to a unipolar obstruction.
}
\setlength{\tabcolsep}{5pt}
\renewcommand{\arraystretch}{1.18}
\footnotesize
\begin{tabular}{lllccccc}
\toprule
$G$ & $\hat{\mathbf n}$ & $G_s(\hat{\mathbf n})$ & $(E_1,E_2)$ &
$\hat\kappa^{(-)}$ & $j^s_\perp$ & W$_\perp$ & SSM \\
\midrule

${}^24/{}^1m{}^1m_x{}^2m_d$ & [001] &
${}^24{}^1m_x{}^2m_d$ & $(E_x,E_y)$ &
$\begin{pmatrix}
0 & a\\
a & 0\\
\mathbf{0} & \mathbf{0}
\end{pmatrix}$ &
$0$ & No & No \\

& [100] ($\perp {}^1m$) &
${}^1m{}^1m{}^12$ & $(E_y,E_z)$ &
$\begin{pmatrix}
\mathbf{0} & \mathbf{0}\\
0 & b\\
c & 0
\end{pmatrix}$ &
$0$ & No & Yes \\

& [110] $\parallel\hat{x}'\perp {}^2m$ &
${}^1m_z{}^2m_{y'}{}^22$ & $(E_{y'},E_z)$ &
$\begin{pmatrix}
\mathbf{0} & \boldsymbol{d}\\
0 & 0\\
0 & 0
\end{pmatrix}$ &
$p\,E_{y'}$ & Yes & No \\

\midrule

${}^24/{}^1m$ & [001] &
${}^24$ & $(E_x,E_y)$ &
$\begin{pmatrix}
f & a\\
a & -f\\
\mathbf{0} & \mathbf{0}
\end{pmatrix}$ &
$0$ & No & No \\

& [100] (generic $\perp {}^24$) &
${}^1m$ & $(E_y,E_z)$ &
$\begin{pmatrix}
\mathbf{0} & \boldsymbol{d}\\
0 & b\\
c & 0
\end{pmatrix}$ &
$p\,E_y$ & Yes & Yes \\

\midrule

${}^22/{}^2m$ & [010] $\parallel {}^22$ &
${}^22$ & $(E_x,E_z)$ &
$\begin{pmatrix}
0 & 0\\
\boldsymbol{a} & \boldsymbol{b}\\
0 & 0
\end{pmatrix}$ &
$p\,E_x+q\,E_z$ & Yes & No \\

& [100] (generic $\perp {}^22$) &
${}^2m$ & $(E_y,E_z)$ &
$\begin{pmatrix}
\mathbf{0} & \boldsymbol{c}\\
d & 0\\
0 & f
\end{pmatrix}$ &
$r\,E_y$ & Yes & No \\

\midrule

${}^2m{}^2m{}^1m_y$ & [010] ($\perp {}^1m$) &
${}^2m{}^2m{}^12$ & $(E_x,E_z)$ &
$\begin{pmatrix}
a & 0\\
\mathbf{0} & \mathbf{0}\\
0 & b
\end{pmatrix}$ &
$0$ & No & No \\

& [100] ($\perp {}^2m$) &
${}^1m_y{}^2m_z{}^22$ & $(E_y,E_z)$ &
$\begin{pmatrix}
\boldsymbol{c} & \mathbf{0}\\
0 & 0\\
0 & 0
\end{pmatrix}$ &
$p\,E_z$ & Yes & No \\

\midrule

${}^26/{}^2m$ & [0001] &
${}^26$ & $(E_x,E_y)$ &
$\begin{pmatrix}
0 & 0\\
0 & 0\\
\mathbf{0}^{\dagger} & \mathbf{0}^{\dagger}
\end{pmatrix}$ &
$0$ & No$^{\dagger}$ & No \\

${}^26/{}^2m{}^1m{}^2m$ & [0001] &
${}^26{}^1m_x{}^2m_y$ & $(E_x,E_y)$ &
$\begin{pmatrix}
0 & 0\\
0 & 0\\
\mathbf{0}^{\dagger} & \mathbf{0}
\end{pmatrix}$ &
$0$ & No$^{\dagger}$ & No \\

\midrule

\begin{tabular}{@{}l@{}}
${}^14/{}^1m{}^2m{}^2m$ \\
${}^16/{}^1m{}^2m{}^2m$
\end{tabular} &
\begin{tabular}{@{}l@{}}
$[001]$ \\
$[0001]$
\end{tabular} &
\begin{tabular}{@{}l@{}}
${}^14{}^2m{}^2m$ \\
${}^16{}^2m{}^2m$
\end{tabular} &
\begin{tabular}{@{}l@{}}
$(E_x,E_y)$ \\
$(E_x,E_y)$
\end{tabular} &
$\begin{pmatrix}
a & 0\\
0 & a\\
\mathbf{0} & \mathbf{0}
\end{pmatrix}$ &
$0$ & No & No \\

\begin{tabular}{@{}l@{}}
${}^1\bar3{}^2m$ \\
${}^1m_z{}^1\bar3_{[111]}{}^2m_d$
\end{tabular} &
\begin{tabular}{@{}l@{}}
$[0001]$ \\
$[111]$
\end{tabular} &
\begin{tabular}{@{}l@{}}
${}^13{}^2m_y$ \\
${}^13{}^2m_{y'}$
\end{tabular} &
\begin{tabular}{@{}l@{}}
$(E_x,E_y)$ \\
$(E_{x'},E_{y'})$
\end{tabular} &
$\begin{pmatrix}
a & 0\\
0 & a\\
\mathbf{0}^{\dagger} & \mathbf{0}
\end{pmatrix}$ &
$0$ & No$^{\dagger}$ & No \\

\bottomrule
\end{tabular}
\end{table*}

The responses listed in Table \ref{tab:atlas} provide clear symmetry design rules for electrically switchable altermagnetic spin-splitter layers. Note that high-symmetry surfaces are typically the most suitable for implementation in epitaxial heterostructures. We find that most antisymmetry point groups and surface orientations, with the exception of the (0001) surface for two $g$-wave groups containing the ${}^26$ axis, permit deterministic switching of at least one component of $\Lvec$ by staggered torque polarizations; the tensor $\hat\kappa^{(-)}$ indicates the required orientations of the electric field. Further, switching of the perpendicular component $L_\perp=\Lvec\cdot\nhat$ is allowed (W$_\perp=\mathrm{Yes}$ in Table \ref{tab:atlas}) if the SAPG contains no rotations about $\nhat$ other than ${}^22$. A vertical mirror fixes the direction of the electric field that can generate out-of-plane $\mathbf{p}^{(-)}_\mu\cdot\nhat$ to be normal to a ${}^1m$ mirror or within a ${}^2m$ antimirror.

Taking the (001) surface of a ${}^24/{}^1m{}^1m_x{}^2m_d$ altermagnet as an example, we see that symmetry permits deterministic switching of $L_x$ by $E_y$, as well as $L_y$ by $E_x$, but $L_z$ is not switchable (hence the W$_\perp=\textrm{No}$ notation). We also see, as expected, that $j^s_\perp$ vanishes by symmetry for some of the surfaces even in $d$-wave altermagnets, ruling out their use as transverse spin current sources via the spin-splitter mechanism.

The most promising surfaces for applications are those that permit both perpendicular switching and $j^s_\perp$. One such surface is $\nhat=[110]$ for the ${}^24/{}^1m{}^1m_x{}^2m_d$ bulk antisymmetry point group, which describes altermagnetic rutiles \cite{Smejkal1} and possibly \cite{thapa2026altermagnetismvanadiumoxychalcogenideslost} some inverse-Lieb-lattice oxychalcogenides \cite{jiang_discovery_2025,zhang_crystal-symmetry-paired_2025,chang2025inverseliebmaterialsaltermagnetism}. Note that in the conventional crystallographic setting used for rutiles (which is rotated by $45^\circ$ compared to the setting used in Table \ref{tab:atlas}), this surface is indexed as [100]. Another relatively high-symmetry surface allowing both perpendicular switching and spin-splitter response is available in the orthorhombic ${}^2m{}^2m{}^1m$ spin group by terminating the crystal along one of the ${}^2m$ antisymmetric mirror planes. This case may be realizable in altermagnets with the TiNiSi-type structure, such as WFeB \cite{gamage2026metallicdwavealtermagnetismwfeb}. Note that both of these surfaces forbid SSM. This means that the two magnetic sublattices remain symmetry-equivalent at the interface, forbidding the nonrelativistic staggered torque-polarization response via unequal absorption of the incident spin current by the two sublattices. An inspection of Table \ref{tab:atlas} shows that most high-symmetry surfaces forbid SSM and hence require interfacial spin-orbit coupling to generate staggered torque polarizations.

Among the cases permitting both perpendicular switching and $j^s_\perp$ in Table \ref{tab:atlas}, all except $G_s(\nhat)={}^22$ require orthogonal electric field directions for perpendicular switching and $j^s_\perp$ generation. This orthogonality is enforced by a mirror plane in achiral SAPGs because the transformation of $p^{(-)}_\perp$ and $j^s_\perp$ under a vertical mirror differs by a sign: the spin index of the former is axial-vector and that of the latter is scalar, while both are sublattice-odd. Moreover, in both cases considered in the previous paragraph, the field orientation driving $j^s_\perp$ generates no staggered torque polarization in  any direction. In device applications, the orthogonality of the writing and operating current paths can minimize interference between the initialization of the N\'eel order and operation of the spin current source.

\paragraph*{Role of the assumptions.}
The entries in Table~\ref{tab:atlas} rely on the assumptions stated in Sec.~\ref{sec:setup}.
In specific heterostructures, the SAPG may be lowered by adjacent layers, growth conditions, or strain \cite{Belashchenko2025}. Such symmetry lowering can permit additional responses, which may be determined from the corresponding subgroup of the SAPG. The responses listed in Table \ref{tab:atlas} represent the minimal set allowed by symmetry for a given surface orientation in the absence of such additional symmetry breaking. We return to this point below in connection with the PtTe$_2$/CrSb (0001) heterostructure studied in Ref. \cite{Li2026}.

The unique bipartite assignment of magnetic atoms to sublattices is preserved for an ensemble of coherently oriented crystalline regions (e.g., grains) related only by translations. If such assignment is not possible in the entire sample, the restrictions on staggered torque polarizabilities obtained from Eq. (\ref{eq:symmetry_condition}) still apply within a subsystem satisfying this assumption, such as within a set of equivalent crystallographic twin domains.

The assumption of inequivalent top and bottom interfaces is important. Indeed, if the two interfaces are symmetry-equivalent, the relevant antisymmetry point group for the film is not the SAPG, which fixes $\nhat$, but a larger group that fixes $\nhat$ only up to a sign. Under the stated assumptions, that group always retains inversion, forbidding all macroscopic current-induced torques linear in $\Evec$ and averaged over the whole film. This means that the contributions to $\hat\kappa^{(-)}$ from symmetry-equivalent top and bottom interfaces cancel each other out.

\section{General angular dependence of the torques}
\label{sec:higher-order}

The general dependence of the torques on the orientation of $\Lvec$ may be represented as an expansion in symmetry-allowed vector spherical harmonics \cite{Belashchenko2020}, in which Eq. (\ref{eq:torques}) is the lowest-order term with $l=1$. Higher-order terms are described by torque multipoles, whose responses to electric field are described by tensors of higher ranks than $\hat\kappa^{(-)}$. Even if some components of $\hat\kappa^{(-)}$ vanish, deterministic switching may still be allowed via higher-order torque harmonics. Whether bistable, bipolar switching considered here is still forbidden, for given directions of $\Evec$ and $\Lvec$, with the general symmetry-allowed angular dependence of the torques, may be established using the following criterion.

Consider a subgroup $G_s(\nhat,\hat\Evec)\leq G_s(\nhat)$ of the SAPG that preserves the axis of $\Evec$ up to a sign. Its only nontrivial spatial operations can be a two-fold rotation about $\nhat$ and vertical mirror planes parallel and perpendicular to $\Evec$. Under $G_s(\nhat,\hat\Evec)$, $\Evec$ transforms according to a real one-dimensional irreducible representation; in the notation of Eq. (\ref{eq:symmetry_condition}), each $g\in G_s(\nhat,\hat\Evec)$ transforms $\Evec$ into $R_g\Evec=e_g\Evec$, where $e_g=\pm1$.

For bistable, bipolar switching, we require that there are two states, $\Lvec$ and $-\Lvec$, that are initialized, respectively, by $\Evec$ and $-\Evec$. Initialization of a unique domain by $\Evec$ requires that the orbit of $\Lvec$ under $G_s(\nhat,\hat\Evec)$ has no elements other than $\Lvec$ and $-\Lvec$. This means $\Lvec$ must transform under a real one-dimensional irreducible representation of $G_s(\nhat,\hat\Evec)$, and we can write 
$\eta_g(\det R_g) R_g\Lvec=s_g\Lvec$ with $s_g=\pm1$.
Furthermore, $\Evec$ and $\Lvec$ must transform under the same irreducible representation. Indeed, $e_g=-s_g=1$ implies that $g$ maps $\Lvec\leftrightarrow-\Lvec$ while $\Evec$ remains unchanged; any switching then must be a toggle. Conversely, $e_g=-s_g=-1$ means $g$ reverses $\Evec$ while leaving $\Lvec$ unchanged; this means that initialization cannot be bipolar. We will refer to the former as a toggle obstruction, and to the latter as a unipolar obstruction. A toggle obstruction is an unbroken switching symmetry in the sense of Ref. \cite{Fu2026}. 

If $\Evec$ and $\Lvec$ transform under the same nontrivial irreducible representation of $G_s(\nhat,\hat\Evec)$, then not only is there no obstruction to switching, but any switching must be strictly bipolar, i.e., if $\Evec$ initializes $\Lvec$, then $-\Evec$ initializes $-\Lvec$ via a symmetry-related process.

All possible constraints imposed by SAPG elements are listed in Table \ref{tab:switching_obstructions}. For example, a ${}^22$ axis reverses both in-plane $\Evec$ and $\Lvec\parallel\nhat$ but leaves in-plane $\Lvec\perp\nhat$ unchanged; this implies that ${}^22$ creates a unipolar obstruction (U) for the switching of $\Lvec\perp\nhat$, but switching of $\Lvec\parallel\nhat$ can only be strictly bipolar (B). Note that pure antisymmetry ${}^21$, which does not occur at altermagnetic surfaces but is retained at any rough surface of an antiferromagnet with a bulk antitranslation, forbids bipolar switching in any geometry.

\begin{table}[htb]
\centering
\caption{\label{tab:switching_obstructions}
Constraints on the switching of a time-reversed domain pair ($\Lvec\leftrightarrow-\Lvec$) imposed by elements of the SAPG. T (toggle obstruction): deterministic initialization is forbidden; any reversal must be a toggle. U (unipolar obstruction): any selection of $\Lvec$ by $\Evec$ is symmetry-related to the selection of the same $\Lvec$ by $-\Evec$. B (strict bipolarity): any selection of $\Lvec$ by $\Evec$ is symmetry-related to the selection of $-\Lvec$ by $-\Evec$. All cases hold for the general, symmetry-allowed angular dependence of the torque, to all orders in $\Evec$.
${}^12$ is contained in ${}^14$, ${}^24$, and ${}^16$; ${}^22$ is contained in ${}^26$.
}
\setlength{\tabcolsep}{9pt}
\renewcommand{\arraystretch}{1.25}

\begin{tabular}{@{}cccc@{}}
\toprule
$g\in G_s(\nhat)$ & $\Evec$ & $\Lvec\parallel\nhat$ &
$\Lvec\perp\nhat$ \\
\midrule
${}^21$ & Any & T & T \\
${}^12$ & Any & U & B \\
${}^22$ & Any & B & U \\
\midrule
${}^1m$ & $\parallel{}^1m$ & T & T: $\Lvec\parallel\Evec$ \\
${}^1m$ & $\perp{}^1m$ & B & U: $\Lvec\parallel\Evec$; B: $\Lvec\perp\Evec$ \\
${}^2m$ & $\parallel{}^2m$ & --- & T: $\Lvec\perp\Evec$ \\
${}^2m$ & $\perp{}^2m$ & U & U: $\Lvec\perp\Evec$; B: $\Lvec\parallel\Evec$ \\
\bottomrule
\end{tabular}

\end{table}

We now apply the rules of Table~\ref{tab:switching_obstructions} to the SAPGs in Table~\ref{tab:atlas} to determine which zero entries of $\hat\kappa^{(-)}$ are fully protected by a toggle or unipolar obstruction for $\Evec$ along the corresponding principal direction. Zero entries without such protection, and the corresponding $W_\perp=\text{No}$ entries, are marked with a dagger. Daggers occur only for $\Lvec\parallel\nhat$ in SAPGs that contain a threefold axis but no ${}^12$ axis, which happens in some $g$-wave and $i$-wave cases. Every undaggered zero entry is enforced by a twofold axis or a mirror mapping the axes of $\Lvec$ and $\Evec$ onto themselves and thereby also imposing a toggle or unipolar obstruction. In contrast, a ${}^13$ axis by itself forces no entry of the in-plane rows of $\hat\kappa^{(-)}$ to vanish, but it zeroes out the out-of-plane row without imposing a general obstruction. Note that obstructions imposed by mirrors apply only for $\Evec$ parallel or perpendicular to the mirror plane; for intermediate directions of $\Evec$, switching of additional components of $\Lvec$ may be allowed unless it is obstructed by a twofold axis.

Consider a perpendicular switching configuration with $\Lvec\parallel\nhat$. Because every SAPG
operation leaves $\nhat$ unchanged, it either leaves the $\pm\Lvec$ domains unchanged or
interchanges them. Therefore, at any order in the torque angular harmonics, any scalar bias between
them induced linearly by $\Evec$ transforms under the SAPG like $p^{(-)}_\perp$ and vanishes
whenever $p^{(-)}_\perp$ does. Such biases include the energy difference produced by a staggered FL
torque and the difference in effective damping produced by a staggered DL torque. Thus, the
vanishing elements of $\hat\kappa^{(-)}$ corresponding to $p^{(-)}_\perp$ that have no associated
toggle or unipolar obstruction, marked by a dagger in Table~\ref{tab:atlas}, still disable the usual
mechanisms in which a current-induced torque either exerts a uniform pressure linear in $\Evec$ on the domain walls or destabilizes one of the two domains.

\section{Illustrative model of the staggered Edelstein effect}

To understand how symmetry-allowed staggered torques can generically emerge in a band model, let us consider a minimal $k\cdot p$ Hamiltonian for the surface bands of a tetragonal altermagnet with bulk antisymmetry group ${}^24/{}^1m{}^1m{}^2m$ and a surface normal along [001], with one effective $s$-like orbital per magnetic site. It is instructive to introduce a structural control parameter $\lambda$ tuning the system continuously from a conventional bipartite antiferromagnet with unit-cell doubling at $\lambda=0$ to an altermagnet at $\lambda=1$, so that the antisymmetry operation ${}^21$ is restored at $\lambda=0$. 
Restricting to in-plane momenta $\mathbf{k}_\parallel=(k_x,k_y)$, the generic nonrelativistic $k\cdot p$ Hamiltonian up to quadratic terms is \cite{Roig2024}:
\begin{align}
    H_0(\mathbf{k})  =  \frac{J}{2} \tau_z \hat\Lvec\cdot\boldsymbol{\sigma}
    +t k_\parallel^2 + (t_0+t_x k_\parallel^2)\tau_x + \lambda t' \tau_z (k^2_y-k^2_x) 
\label{H0}
\end{align}
where $\boldsymbol{\sigma}$ and $\boldsymbol{\tau}$ are Pauli matrices in spin and sublattice space, respectively. At $k_\parallel=0$ there are two doublets at $\epsilon_\pm=\pm\Delta/2$, where  $\Delta=\epsilon_+-\epsilon_-=\sqrt{J^2+4t_0^2}$.
We assume that we are dealing with conduction bands and only the lower doublet crosses the Fermi level $E_F$, measured from $\epsilon_-$, such that $E_F\ll\Delta$.

On top of $H_0$, we add spin-orbit coupling terms allowed by the symmetry of the film with the (oriented) surface normal $\hat{\mathbf n}=[001]$. Imposing invariance with respect to the ${}^24{}^1m{}^2m$ SAPG (at $\lambda\neq0$), and retaining only time-reversal-even terms, we obtain three symmetry-allowed spin-orbit couplings linear in $\mathbf{k}$:
\begin{align}
H_{\mathrm{SOC}}(\kvec)=\boldsymbol{b}(\mathbf{k})\cdot\boldsymbol{\sigma}&=
(\alpha_0 + \alpha_x \tau_x) (k_y\sigma_x-k_x\sigma_y)\nonumber\\
&+\lambda\beta_z (k_y\sigma_x+k_x\sigma_y)\tau_z ,
\label{HSOC}
\end{align}
which include two sublattice-even Rashba couplings $\alpha_0$, $\alpha_x$ and one sublattice-odd Dresselhaus-like coupling $\beta_z$. Note that the effective momentum-dependent spin-orbit field $\boldsymbol{b}(\mathbf{k})$ is an operator in sublattice space. Because the Dresselhaus term is odd under antisymmetry, it vanishes at $\lambda=0$ where ${}^21$ is restored.

Current-induced shift of the Fermi surface generates FL torques through the Edelstein mechanism. Assuming constant transport relaxation time $\tau_{tr}$, weak altermagnetic band anisotropy, $|\lambda t'/t|\ll1$, to first order in $\beta_z$ the Dresselhaus term generates a staggered FL torque polarization $\mathbf{p}^{(-)}_\mathrm{FL}(\mathbf{E})=a(E_y,E_x,0)$ per unit area, where
$a=-(e/\hbar) \lambda\beta_z \tau_{tr}(J/\Delta) n_-$, and $n_-$ is the areal carrier density for the occupied lower doublet. The tensor structure of this response conforms with the symmetry-allowed $\hat\kappa^{(-)}$ tensor listed in Table \ref{tab:atlas} for the ${}^24{}^1m{}^2m$ SAPG. While Rashba-like terms primarily generate a sublattice-even FL torque polarization $\mathbf{p}^{(+)}_\mathrm{FL}\propto\hat{\mathbf z}\times\mathbf{E}$, they also contribute nonperturbatively to $\mathbf{p}^{(-)}_\mathrm{FL}$ by gapping the altermagnetic nodal lines at $|k_x|=|k_y|$.

This staggered Edelstein effect (interfacial NSOT) generated by interfacial spin-orbit coupling provides a microscopic realization of the symmetry-allowed $\hat\kappa^{(-)}_\mathrm{FL}$ response. It is generally applicable because the symmetry-allowed $\kappa^{(-)}_{ij}$ response components and the sublattice-odd interfacial spin-orbit couplings $\tau_z \sigma_ik_j$ transform identically under the SAPG. For example, the (110) surface of a ${}^24/{}^1m{}^1m_x{}^2m_d$ altermagnet supports $H_\mathrm{SOC}(\kvec)=\lambda\beta_\perp\tau_z\sigma_{x'}k_z$, which corresponds to the single nonzero component $\kappa^{(-)}_{x'z}=d$ appearing in Table \ref{tab:atlas}. In the presence of $E_z$, this term generates a staggered spin accumulation (and hence a staggered FL torque polarization) parallel to the surface normal $\hat x'\parallel[110]$. Staggered responses of the same symmetry are also allowed in the DL channel.

Under our statistical averaging assumptions, the staggered torque polarizations $\mathbf{p}^{(-)}_\mu(\mathbf{E})$ vanish both in the bulk \emph{and} at the surface of a cell-doubling antiferromagnet ($\lambda=0$), because they are forbidden by the ${}^21$ antisymmetry operation associated with the cell-doubling magnetic translation. Current-induced bipolar switching is generally forbidden in this case by a toggle obstruction that is valid to all orders in angular torque harmonics.

\section{Symmetry of switching in C\lowercase{r}S\lowercase{b} films}

Let us apply the SAPG method to analyze the symmetry of CrSb film configurations used in Ref. \cite{Zhou2025}. In all three configurations, the hexagonal axis $\hat{\mathbf{c}}=[0001]$ lies in the film plane. Configuration I is a CrSb ($11\bar20$) film whose SAPG is ${}^2m{}^2m{}^12$; Table \ref{tab:atlas} shows that both in-plane components of $\Lvec$ can be switched by parallel components of $\Evec$ in this SAPG. Configuration II is a CrSb ($1\bar100$) film with a different SAPG ${}^1m{}^2m{}^22$, which only allows electrical switching of the perpendicular $\Lvec\parallel\nhat$ component by an in-plane field $\Evec\parallel\hat{\mathbf{c}}\times\nhat$. Finally, Configuration III is CrSb ($11\bar20$) but with an additional monoclinic distortion that removes the vertical antimirror plane parallel to the hexagonal axis while preserving the (0001) antimirror. This leaves us with $G_s(\nhat)={}^2m$, which, according to Table \ref{tab:atlas}, allows three independent components of $\hat\kappa^{(-)}$: two that are already allowed for Configuration I, and the third one allows switching of $\Lvec\parallel\nhat$ by $\Evec\parallel\hat{\mathbf{c}}\times\nhat$. Because  $\Lvec\parallel\hat{\mathbf{c}}$ in CrSb, it can be switched by $\Evec\parallel\hat{\mathbf{c}}$ in Configurations I and III but not in Configuration II; the latter case is also forbidden to all orders by a toggle obstruction enforced by a ${}^1m$ mirror plane containing $\hat{\mathbf{c}}$ \cite{Fu2026}. Switching was indeed observed in Configuration III, which also allows an anomalous Hall effect with the Hall vector component along $\nhat$ \cite{Zhou2025}. Note that none of the three configurations allow SSM.

Zhou \emph{et al.} \cite{Zhou2025} ascribed the deterministic driving force to an effective coupling $\Delta E_\mathrm{DM}\propto-\Lvec\cdot(\mathbf{p}\times\mathbf{D})$, where $\mathbf{p}$ is the current-induced spin polarization injected through the interface, and $\mathbf{D}$ is the Dzyaloshinskii-Moriya vector. In the notation of the present paper, $\mathbf{p}$ is symmetry-equivalent to the sublattice-even $\mathbf{p}^{(+)}_\mathrm{FL}=\hat\kappa^{(+)}_\mathrm{FL}\Evec$, and $\Delta E_\mathrm{DM}$ corresponds to  $\hat\kappa^{(-)}_\mathrm{FL}\propto\mathbf{D}\times\hat\kappa^{(+)}_\mathrm{FL}$. This term is compatible with the SAPG because $\mathbf{D}$ is axial and sublattice-odd. A similar term is allowed in the DL channel. However, these terms do not generally span the full symmetry-allowed structure of the $\hat\kappa^{(-)}_\mu$ tensors. For example, the ${}^24$ operation enforces $\mathbf{D}=0$, and yet, as seen from Table \ref{tab:atlas}, the $\hat\kappa^{(-)}$ tensors do not vanish in the SAPGs containing this operation.

Note that, in contrast to the interfacial NSOT mechanism appearing already in linear order in spin-orbit coupling, the mechanism mediated by Dzyaloshinskii-Moriya interaction \cite{Zhou2025} requires spin-orbit coupling twice: first to generate a sublattice-even spin polarization and then to convert it into staggered torque.

An example of additional symmetry lowering by an adjacent layer is provided by the epitaxial PtTe$_2$/CrSb (0001) heterostructure studied in Ref. \cite{Li2026}. Its interfacial SAPG is ${}^13{}^1m$ \cite{Li2026}, which is lower than ${}^26{}^1m{}^2m$ of the free CrSb (0001) surface because trigonal PtTe$_2$ breaks the ${}^2m$ antimirror. Because the ${}^13{}^1m$ SAPG retains no sublattice-swapping operations, it allows SSM which is exchange coupled to $\Lvec$ \cite{Andreev1996,Belashchenko2010}. SAPG invariance (\ref{eq:symmetry_condition}) only allows $\mathbf{p}^{(-)}_\mu\propto\nhat\times\Evec$, which, thanks to the presence of SSM, appears in tandem with the conventional uniform $\mathbf{p}^{(+)}_\mu\propto\nhat\times\Evec$ as in ferromagnet/heavy-metal bilayers \cite{Manchon-RMP}. Because these $\mathbf{p}^{(-)}_\mu$ are in-plane, they alone cannot switch perpendicular $\Lvec$ in the PtTe$_2$/CrSb (0001) bilayer with this SAPG; an assisting in-plane magnetic field was used in Ref. \cite{Li2026}. Note, however, that switching of perpendicular $\Lvec$ by higher-order torques is not generally prohibited even for the full ${}^26{}^1m{}^2m$ SAPG of CrSb (0001) if the current does not flow within a ${}^1m$ mirror plane, as indicated by a dagger in Table \ref{tab:atlas}. The lowest staggered torque harmonics in CrSb (0001) are the sublattice-odd counterparts of so-called $3m$ torques with $l=2$ \cite{Liu2021,Ovalle2023,Fang2025}, and they are also allowed in the PtTe$_2$/CrSb (0001) system in addition to the in-plane staggered $\mathbf{p}^{(-)}_\mu$. Field-free switching of perpendicular magnetization was observed in CoPt/CuPt(111) bilayers \cite{Liu2021}, which have the same harmonics allowed in the uniform channel as PtTe$_2$/CrSb (0001) has in both uniform and staggered channels. Switching of $L_\perp$ enabled by higher-order torques cannot rely on a bias between $\pm\Lvec$ linear in $\Evec$ (Sec.~\ref{sec:higher-order}) and must therefore involve different mechanisms, such as those studied for ferromagnets \cite{Fang2025,Fokkens2026}.

\section{Conclusions}

Deterministic electrical 180$^\circ$ switchability of the N\'eel vector in a film of a bulk-centrosymmetric altermagnet is determined by its surface antisymmetry point group (SAPG). The allowed staggered torque-polarizability tensors are classified in Table \ref{tab:atlas}, and general obstructions to switching that hold to all orders in torque angular harmonics are listed in Table \ref{tab:switching_obstructions}. There is a particularly simple criterion for the perpendicular component $L_\perp$: its switching by current-induced staggered torque polarizations is allowed if the SAPG contains no rotations other than a sublattice-interchanging twofold axis. Moreover, a sublattice-preserving twofold axis forbids bipolar switching of $L_\perp$ to all orders in torque angular harmonics and $\Evec$. Symmetry-allowed staggered fieldlike torques generically appear due to the Edelstein effect generated by the sublattice-odd interfacial spin-orbit coupling. Several surfaces permit both deterministic electrical writing of a perpendicular N\'eel vector and generation of a perpendicular spin current $j^s_\perp$ via the nonrelativistic spin-splitter effect. If such a surface retains a mirror plane in its SAPG, the writing and operating electric fields are orthogonal.
These design rules enable the implementation of switchable altermagnetic layers both as information storage elements and as spin current sources in spin-torque devices.

\begin{acknowledgments}
I thank Alexey Kovalev and Fei Xue for useful discussions. This work was supported by the U.S. Department of Energy (DOE) Established Program to Stimulate Competitive Research (EPSCoR) through Grant No. DE-SC0024284. I used OpenAI GPT 5.2--6 and Anthropic Claude Opus 5.5 models interactively to explore and refine theoretical arguments and derivations, and to organize and edit the manuscript. I have critically evaluated and verified all model output used in this work.
\end{acknowledgments}

%


\end{document}